\documentclass[aps,prl,twocolumn,showpacs,superscriptaddress,groupedaddress]{revtex4}

\usepackage{amssymb}
\usepackage{color,graphicx}
\usepackage{amsmath}
\usepackage{amsbsy}
\usepackage{amsthm}
\usepackage{bbm}
\usepackage{bm}
\usepackage{epsfig}
\usepackage{lscape}
\usepackage{float}
\usepackage{subfigure}

\begin{document}

\title{Testing the quantum superposition principle in the frequency domain}

\author{M. Bahrami}
\email{mbahrami@ictp.it}
\affiliation{Department of Physics, University of Trieste, Strada Costiera 11, 34014 Trieste, Italy}
\affiliation{%Condense Matter and Statistical Physics, 
The Abdus Salam ICTP, Strada Costiera 11, 34151 Trieste, Italy}

\author{A. Bassi}
\email{bassi@ts.infn.it}
\affiliation{Department of Physics, University of Trieste, Strada Costiera 11, 34014 Trieste, Italy}
\affiliation{Istituto
Nazionale di Fisica Nucleare, Trieste Section, Via Valerio 2, 34127 Trieste,
Italy}

\author{H. Ulbricht}
\email{h.ulbricht@soton.ac.uk}
\affiliation{School of Physics and Astronomy, University of Southampton, Southampton SO17 1BJ, United Kingdom}

\date{\today}

\begin{abstract}
{New technological developments allow to explore the quantum properties of very complex systems, bringing the question of whether also macroscopic systems share such features, within experimental reach. The interest in this question is increased by the fact that, on the theory side, many suggest that the quantum superposition principle is not exact, departures from it being the larger, the more macroscopic the system. 
% Models of spontaneous wave function collapse have been engineered to include such a possibility, by adding nonlinear stochastic terms to the Schr\"odinger equation. As such, they make predictions which differ from standard quantum predictions. 
Here we propose a novel way to test the possible violation of the superposition principle, by analyzing its effect on the spectral properties of a generic two-level system. We will show that spectral lines shapes are modified, if the superposition principle is violated, and we quantify the magnitude of the violation. We show how this effect can be distinguished from that of standard environmental noises. We argue that accurate enough spectroscopic experiments are within reach, with current technology.}
\end{abstract}

\pacs{}

\maketitle

{\it Introduction.} 
-- Quantum theory is an extremely successful theory of microscopic phenomena. Yet, still an open question is if and how it applies also to macroscopic systems. Indeed it would be absolutely fascinating to see quantum mechanical behavior for a macroscopic object. But until today macroscopic systems are always found to behave according to classical mechanics and the open question is exactly if they can show quantum behaviour or if the quantum superposition principle is violated at some level. Here we show how such violations would modify spectral lineshapes of atoms and molecules, and we propose a new class of spectroscopic experiments to test such modifications. These violations have a universal character, which cannot be controlled by experimental parameters. We will show how this specific signature can be used to distinguish them from decoherence.

In general, when performing a quantum experiment, two different limitations can be identified. We distinguish the technological and the fundamental limitations. Technological limitations are induced through coupling of the quantum system with the environment. Decoherence theory explains how environmental noise can rapidly destroy quantum coherence before it can be even observed~\cite{Zurek}. For macroscopic systems we can safely expect a plethora of decohering interactions and it will be a technological challenge to isolate large quantum systems from such interactions. On the other hand fundamental limitations are due to intrinsic non-linearity in the quantum dynamics that are induced e.g., by the gravity as in the Di\'{o}si-Penrose model~\cite{dp}, or by a generic stochastic field as in the Continuous Spontaneous Localization model~\cite{Csl,collapse_review1,collapse_review2}. In all such models, Schr\"{o}dinger dynamics is modified by adding non-linear terms, in such a way that quantum coherence is preserved at the microscopic level, while it is destroyed when approaching the macroscopic scale~\cite{new_phys0,new_phys1,new_phys2,new_phys3}. % We note that to model this transition, stochasticity is of fundamental importance to avoid superluminal signalling (see Methods).

Recently, there has been rapid experimental progress in revealing quantum features such as particle-wave duality for large objects with tiny de Broglie wavelength of only a few hundred femtometer. Such objects were successfully decoupled from environmental noises, thus overcoming the technological limit and thereby extending the realm of quantum theory to new regimes~\cite{nano1,nano2,super1,super2,super3,opto1,opto3,macro_mol3,macro_mol2}. This progress not only confirms the predictions of decoherence theory, but also provides the possibility to search for intrinsic non-linearity. 

Current experiments are mainly focused on the preparation of macroscopic systems in a spatial quantum superposition state. Non-linearity would then manifest as loss of visibility in observed inference pattern~\cite{collapse_review2}. Promising schemes to observe such non-linearity are provided by optomechanical techniques~\cite{opto1,opto3} and matter-wave  interferometry~\cite{macro_mol2,macro_mol3}. They have not yet found any evidence for intrinsic non-linearity, however allow to introduce upper bounds on their strength~\cite{macro_mol2}. The definite test of the existence of fundamental limitations to the quantum superposition principle will require quantum interference with single particles of mass $10^6-10^9\,$amu~\cite{opto3,macro_mol3}.  
 
Here, we provide an alternative test of fundamental limitations by measuring light emitted by radiative transitions of excited states of matter. We compute modifications to the standard quantum optical formulation of spontaneous photoemission, assuming quantum linearity is violated. 
% By integrating the Heisenberg equations of motion for the modified dynamics, 
We derive a analytical formula for both shift and broadening of spectral lines, which is applicable to a broad class of systems. We explicitly compute these effects for specific models and then quantify them for relevant physical systems. We show that observation of line broadening induced by intrinsic non-linearity is at the reach of current technology. 

\begin{figure*}[t]
\hspace*{0.1cm}
\subfigure(a){\includegraphics[width=0.45\textwidth]{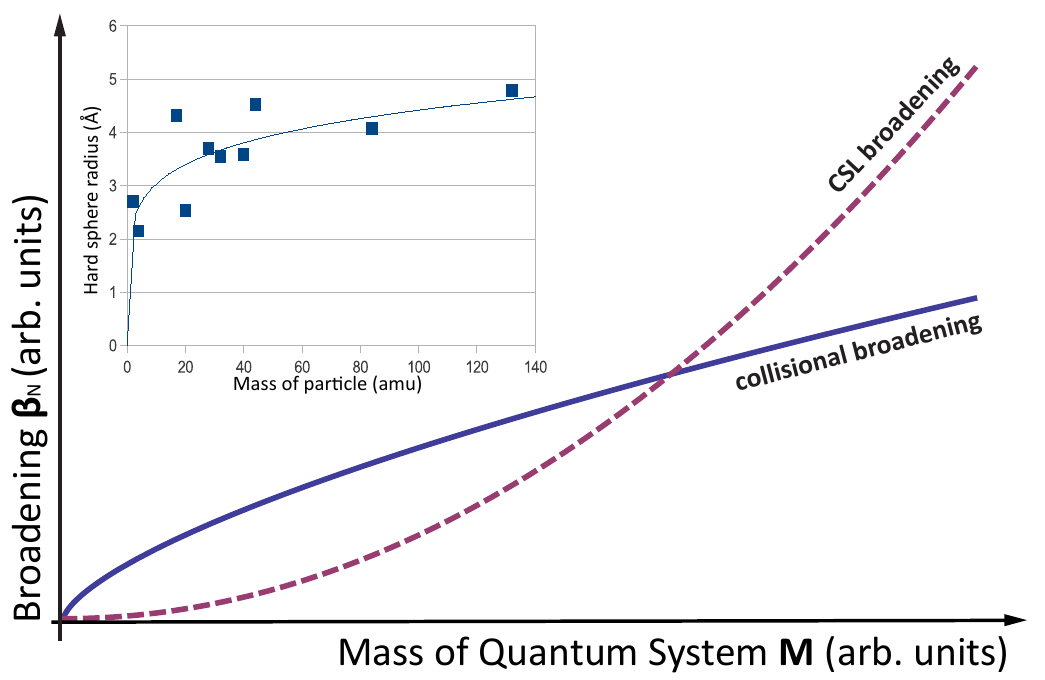}}
\subfigure(b){\includegraphics[width=0.45\textwidth]{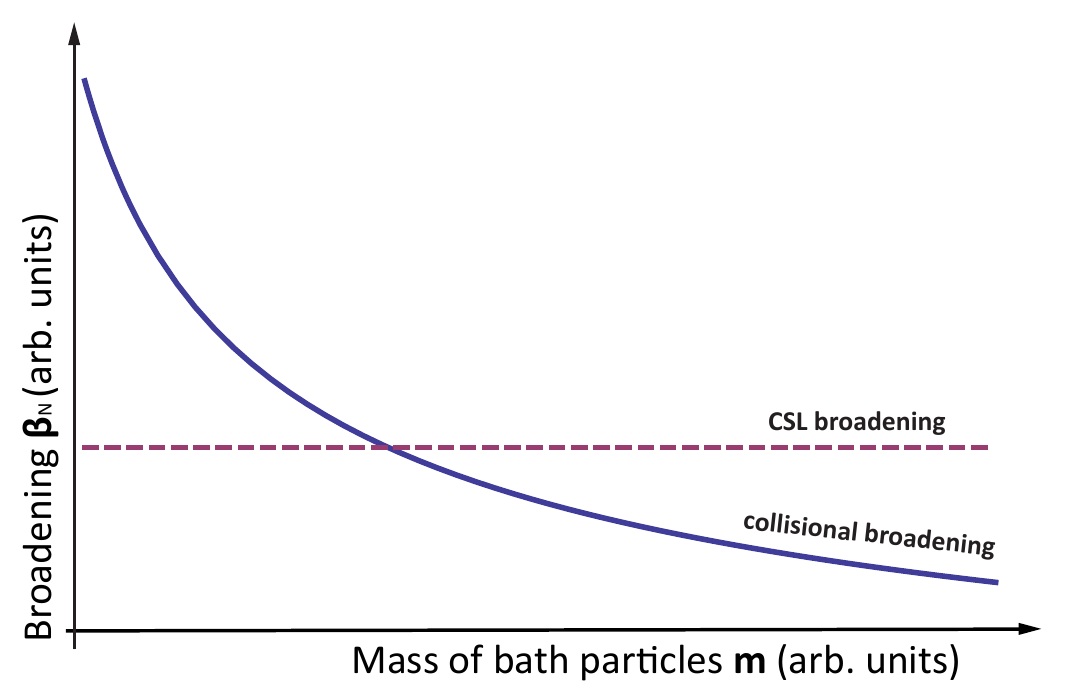}}
\caption{Collapse models predicts that the broadening $\beta_{\text{\tiny N}}$ depends only on the geometry of involved states, and scales quadratically with the mass of the system (see Eq.~\eqref{eq:2}). On the other hand, decoherence broadening in general scales differently with the mass, and moreover depends on the details of the interaction with the environment. In Fig.~(a) we show the different scaling behavior between CSL broadening and decoherence broadening (in the recoil-free regime $M\gg m$) while the environment does not change and the system increases in mass. Assuming the density of the system to remain constant so that $r \propto M^{1/3}$ ($r$ is the radius of the systems and $M$ its mass), then in the Hard sphere limit one has for collisional broadening~\cite{Loudon}: $\beta_{\text{\tiny C}} = 4 d^2 p \sqrt{\pi/m k_{\text{\tiny B}} T} \propto M^{2/3}$ (where $d \sim r$ is the closet distance among the colliding particles, $p$ the pressure of the bath, $T$ its temperature and $m$ the mass of bath particles). In the inset,  the dependence of the experimental Hard sphere radius on mass is shown, for H$_2$, He, NH$_3$, Ne, N$_2$, O$_2$, Ar, CO$_2$, Kr, and Xe~\cite{crc}. The behaviour is qualitatively similar to the oversimplified $M^{1/3}$ dependence we estimated, therefore the mass dependence of $\beta_{\text{\tiny C}}$ is different from the $M^2$ behavior of the CSL broadening.
In Fig. (b) we assume the opposite situation. The system does not change and the bath particle increases in mass (still in the recoil-free regime $M\gg m$). In such a case, CSL broadening remains constant, while decoherence broadening scales as $m^{-1/2}$.}
\label{fig:2}
\end{figure*}

%%%%%%%%%%%%%%%%%%%%%%%%%%%%%%%%%%%%%%%%%%%%%%%%%%%%%%%%%%%%%%%%%%%%%%%%%%%%%%%%%%%%%5
\noindent {\it The frequency shift and broadening.} 
%%%%%%%%%%%%%%%%%%%%%%%%%%%%%%%%%%%%%%%%%%%%%%%%%%%%%%%%%%%%%%%%%%%%%%%%%%%%%%%%%%%%%5
-- Very similar to vacuum fluctuations, intrinsic non-linearity produces two types of radiative corrections in the emitted light spectrum: frequency shift and broadening. Both phenomena appear naturally in QED analysis of the spectrum of spontaneously emitted light. We formulate the intrinsic non-linearity in terms of a stochastic potential added to the Schr\"odinger equation, since observable effects of nonlinearity, at the statistical level, can be reproduced by this stochastic potential~\cite{supp}. The most convenient form of stochastic potential describing the destruction of spatial quantum superpositions is given by: 
\begin{eqnarray}
\label{eq:noise}
\hat{V}_t = - \hbar\,\sqrt{\lambda} \int d^3x\; \hat{L}({\bf x})\, \xi_t({\bf x}),
\end{eqnarray} 
with $\xi_t({\bf x})$ a c-number white-noise field, $\lambda$ the strength of nonlinearity, and $\hat{L}({\bf x})$ suitably-chosen Lindblad operators mimicking the nonlinear (collapse) effect. 

We apply this model to describe the dynamics of a two-level system interacting with photons. The two-level representation of stochastic potential $\hat{V}_t$ is obtained by calculating $\langle\varepsilon_\alpha|\hat{V}_t|\varepsilon_\beta\rangle$ with $\alpha,\beta=1,2$. 
Eigenenergies are real functions in most cases, thus one finds:
\begin{equation}
\label{eq:noise2d}
\hat{V}_t = -\hbar\left(
\sqrt{\lambda_z}\,w_t^{(z)}\,\hat{\sigma}_z
+\sqrt{\lambda_x}\,w_t^{(x)}\,\hat{\sigma}_x
\right)
\end{equation}
where the rates $\lambda_{x,z}$ determine the strength of non-linearity and $w^{(x,z)}_t$ are white noises. Later, we will show that $\lambda_{x,z}$ can be obtained from computing the terms $\mathbb{E}(\langle\varepsilon_{\alpha'}|\hat{V}_{t_1}|\varepsilon_{\beta'}\rangle\langle\varepsilon_\alpha|\hat{V}_{t_2}|\varepsilon_\beta\rangle)$ where $\mathbb{E}(\cdot)$ denotes stochastic averaging.
Accordingly, the dynamic of a two-level system interacting with the stochastic field and also with photons is fully described by the Hamiltonian $\hat{H}=\hat{H}_0+\hat{V}_t$, where $\hat{H}_0$ is the standard quantum Hamiltonian characterizing the interaction between a two-level system and a quantized radiation field, which is, in the dipole approximation, given by~\cite{Loudon}:
$\hat{H}_0=\hbar
\sum_{s,\mathbf{k}} 
\omega\,\hat{a}^\dagger_{s,\mathbf{k}}\hat{a}_{s,\mathbf{k}}
+
(\hbar\omega_{0}/2)\hat{\sigma}_z
+\hbar\omega_{0}  \sum_{s,\mathbf{k}} \left\lbrace
g_{s,\mathbf{k}}\hat{\sigma}_y\,\hat{a}_{s,\mathbf{k}} 
- \text{H.C.}\right\rbrace$,
with $g_{s,\mathbf{k}} = (2\epsilon_0\hbar\omega L^3)^{-1/2}\,
\mathbf{d}_{12}\cdot\mathbf{e}_{s,\mathbf{k}}$. Here the system's Hamiltonian is $(\hbar\omega_0/2)\hat{\sigma}_z$, the position operator is proportional to $\hat{\sigma}_x=\hat{\sigma}_++\hat{\sigma}_-$, and the transition dipole matrix element between two levels is not zero ($\mathbf{d}_{12}=\langle\varepsilon_1|\hat{\mathbf{d}}|\varepsilon_2\rangle\neq0$). %with the dipole operator $\hat{{\bf d}} = \sum_i e_i \hat{{\bf q}}_i$). 
All other terms have the usual meanings. According to the Wiener-Khinchin theorem~\cite{Loudon}, the light spectrum is given by the Fourier transform of the normalized dipole-dipole autocorrelation function.
Measurable quantities are obtained after taking the stochastic averages. 
Therefore, all we need is to solve the corresponding differential equations for dipole-dipole autocorrelation functions. This is a lengthy calculation, fully reported in~\cite{supp}; here, we present only the final result for the stochastic averaged equations that we need to compute the dipole-dipole autocorrelation. They read:
\begin{eqnarray}
\nonumber
\lefteqn{\frac{d}{d\tau}\mathbb{E}(\langle\hat{\sigma}_y(t+\tau)\hat{\sigma}_-(t)\rangle)=
\Omega_{\text{\tiny QED}}\,\mathbb{E}(\langle\hat{\sigma}_x(t+\tau)\hat{\sigma}_-(t)\rangle)}
\\&&
-(\beta_{\text{\tiny QED}}+2\lambda_x+2\lambda_z)
\mathbb{E}(\langle\hat{\sigma}_y(t+\tau)\hat{\sigma}_-(t)\rangle),
\\\nonumber
\lefteqn{\frac{d}{d\tau}\mathbb{E}(\langle\hat{\sigma}_x(t+\tau)\hat{\sigma}_-(t)\rangle)=
-\Omega_{\text{\tiny QED}}\,\mathbb{E}(\langle\hat{\sigma}_y(t+\tau)\hat{\sigma}_-(t)\rangle)}
\\
&&-(\beta_{\text{\tiny QED}}+2\lambda_z)
\mathbb{E}(\langle\hat{\sigma}_x(t+\tau)\hat{\sigma}_-(t)\rangle).
\end{eqnarray}
where $\beta_{\text{\tiny QED}}=\frac{\omega_0^3|\mathbf{d}_{12}|^2}{6\pi\epsilon_0\hbar c^3}
$ is the standard spontaneous photo-emission rate and $\Omega_{\text{\tiny QED}} = \omega_0-\gamma_{\text{\tiny QED}}$ with
$\gamma_{\text{\tiny QED}}=
\frac{\omega_0^2|\mathbf{d}_{12}|^2}{3\epsilon_0\hbar\pi^2c^3}
\left(\mathcal{P}\int_0^\infty \frac{d\omega \,\omega}{\omega-\omega_0}
-\mathcal{P}\int_0^\infty \frac{d\omega \,\omega}{\omega+\omega_0}\right)$ the Lamb shift (where $\mathcal{P}$ is Cauchy principal part).
Accordingly, for the dipole-dipole autocorrelation function we find:
\begin{eqnarray}
\lefteqn{\mathbb{E}(\langle\hat{\sigma}_+(t+\tau)\hat{\sigma}_-(t)\rangle)
=
e^{-(\beta_{\text{\tiny QED}}+\lambda_x+2\lambda_z)\tau}\times}
\\\nonumber&&
\left(
\cos\Omega \tau+i\frac{\Omega_{\text{\tiny QED}}}{\Omega}\sin \Omega\tau
\right)
\mathbb{E}(\langle\hat{\sigma}_+(t)\hat{\sigma}_-(t)\rangle)
\end{eqnarray}
with $\Omega=\sqrt{{\Omega^2_{\text{\tiny QED}}}-\lambda_x^2}$ for $\Omega_{\text{\tiny QED}}>\lambda_x$. For $\Omega_{\text{\tiny QED}}\gg\lambda_x$, which is the case for most cases of experimental interest, we can Taylor-expand $\Omega_{\text{\tiny QED}}/\Omega$ to the first leading term in $\lambda_x/\Omega_{\text{\tiny QED}}$, and we get: $
\mathbb{E}(\langle\hat{\sigma}_+(t+\tau)\hat{\sigma}_-(t)\rangle)
\simeq
e^{-(\beta_{\text{\tiny QED}}+\lambda_x+2\lambda_z)\tau}\,e^{i\Omega\tau}\mathbb{E}(\langle\hat{\sigma}_+(t)\hat{\sigma}_-(t)\rangle)$. 
Using Wiener-Khinchin relation, 
for the spectral density of emitted radiation we obtain:
\begin{eqnarray}
%\nonumber
\mathcal{S}(\omega)&=&
%\frac{1}{\pi}\left(
%\frac{\beta_{\text{\tiny QED}}+\lambda_x+2\lambda_z}{(\beta_{\text{\tiny QED}}+\lambda_x+2\lambda_z)^2+(\omega-\Omega)^2}
%\right) 
%\\
\label{eq:sw} 
% & \equiv &  
\frac{1}{\pi}\left(
\frac{\beta}{\beta^2+(\omega-\Omega)^2}\right)\,,
\end{eqnarray}
which is a Lorentzian distribution with the central frequency $\Omega\simeq \Omega_{\text{\tiny QED}}-\gamma_{\text{\tiny N}}$ where $\gamma_{\text{\tiny N}}=\lambda_x^2/2\omega_0$ is the shift induced by intrinsic non-linearity; and the full width at half-maximum of $2\beta=2\beta_{\text{\tiny QED}}+2\beta_{\text{\tiny N}}$ where $\beta_{\text{\tiny N}}=\lambda_x+2\lambda_z$ is the broadening induced by intrinsic non-linearity. For $\beta_{\text{\tiny N}}=\gamma_{\text{\tiny N}}=0$, the standard quantum optical spectrum is recovered. 
Accordingly, the radiative corrections due to the modified dynamics manifest both as  frequency shift and broadening. We shall see that broadening is much bigger than the shift.

%%%%%%%%%%%%%%%%%%%%%%%%%%%%%%%%%%%%%%%%%%%%%%%%%%%%%%%%%%%%%%%%%%%%%%%%%%%%%%%%%
\noindent{\it Quantifying the shift and the broadening}.
%%%%%%%%%%%%%%%%%%%%%%%%%%%%%%%%%%%%%%%%%%%%%%%%%%%%%%%%%%%%%%%%%%%%%%%%%%%%%%%%%
-- We now obtain the shift and the broadening from two most studied collapse models in the literature: the mass proportional Continuous Spontaneous Localization (CSL) model~\cite{Csl,collapse_review1,collapse_review2} and the Di\'{o}si-Penrose (DP) gravitation model~\cite{dp}. For both models, we have~\footnote{This form of $\hat{L}(\mathbf{x})$ is different from Di\'{o}si's original proposal~\cite{dp}. To avoid the divergence due to the Newton self-energy, Di\'{o}si initially proposed a Lindblad operator with a length cutoff equal to the nuclear size ($10^{-15}-10^{-14}\,$m). Then, it was shown that with this small cutoff, DP model contradicts some known observations~\cite{ggr}. To avoid this problem, Ghirardi, Grassi and Rimini~\cite{ggr} proposed this new form of Lindblad operator whose cutoff is $r_C\simeq10^{-7}\,$m. This adjustment of the model was eventually acknowledged by Di\'{o}si~\cite{braz}.}:
\begin{eqnarray}
\label{eq:Lx}
\hat{L}({\bf x})=
\int d\mathbf{y}\frac{
e^{-\mathbf{|\mathbf{x-y}|}^{2}/2r_{C}^{2}}}
{(\sqrt{2\pi}r_{C})^{3}}
\,\sum_{\mathfrak{j}}\,m_\mathfrak{j}\,\hat{a}_{\mathfrak{j}}^{\dagger}\left(\mathbf{y}\right)\hat{a}_{\mathfrak{j}}\left(\mathbf{y}\right)
\end{eqnarray}
where the sum is over the type of particles (electrons and nucleons, in the nonrelativistic limit), $r_C\simeq10^{-7}\,$m the correlation length, and $\hat{a}_{\mathfrak{j}}\left(\mathbf{y}\right)$ is the annihilation operator of a particle with mass $m_\mathfrak{j}$ at the position $\mathbf{y}$. 
Regarding Eq.~\eqref{eq:noise}, for the CSL model we have $\lambda=\gamma/m_0^2$
(with $m_0=1\,$amu, $\gamma \simeq 10^{-22}\text{cm}^{3}\text{s}^{-1}$~\cite{collapse_review2}) and $\xi_t(\mathbf{x})$ is a white noise field with correlation $\mathbb{E}(\xi_{t_1}(\mathbf{x})\xi_{t_2}(\mathbf{y}))=\delta(t_1-t_2)\delta(\mathbf{x}-\mathbf{y})$.
For the DP model, $\lambda=1$ and $\mathbb{E}(\xi_t(\mathbf{x})\xi_s(\mathbf{y}))=G\,\delta(t-s)/\hbar\,|\mathbf{x}-\mathbf{y}|$ with $G$ the gravitational constant. 

Taking into the account Eq.~\eqref{eq:noise2d} and the correlation properties of the noises, one can show that:
$\mathbb{E}(\langle\varepsilon_{1}|\hat{V}_{t_1}|\varepsilon_{1}\rangle\langle\varepsilon_1|\hat{V}_{t_2}|\varepsilon_1\rangle)
-\mathbb{E}(\langle\varepsilon_{2}|\hat{V}_{t_1}|\varepsilon_{2}\rangle\langle\varepsilon_2|\hat{V}_{t_2}|\varepsilon_2\rangle) \equiv 2\lambda_z\,\hbar^2\,\delta(t_1-t_2)$ and $\mathbb{E}(\langle\varepsilon_{2}|\hat{V}_{t_1}|\varepsilon_{1}\rangle\langle\varepsilon_2|\hat{V}_{t_2}|\varepsilon_1\rangle) \equiv \lambda_x\,\hbar^2\,\delta(t_1-t_2)$.
From these relations, together with Eq.~\eqref{eq:sw}, one can obtain the analytical expression for the broadening $\beta_{\text{\tiny N}}$ and the shift $\gamma_{\text{\tiny N}}$. 
We compute the shift and broadening for the CSL model and DP model when the region in which the amplitude of the eigenstates is different from zero, is smaller than $r_C\simeq10^{-7}$m, which is valid for most cases of interest such as molecules.
After a lengthy computation, reported in~\cite{supp}, we get:
\begin{eqnarray}
\label{eq:2}
\beta_{\text{\tiny N}}&=&\frac{\Lambda}{2r_C^2\,m_0^2}
\left(D^2_{12}+\frac{M}{2}\left(I_2-I_1\right)\right),\\
\label{eq:3}
\gamma_{\text{\tiny N}}&=&\frac{\Lambda^2}{8\omega_0}\left(
\frac{D_{12}}{r_C\,m_0}\right)^4,
\end{eqnarray}
with $I_\alpha=
\langle\varepsilon_\alpha|(\sum_{j=1}^Nm_j\,\hat{\mathbf{q}}_j^2)|\varepsilon_\alpha\rangle$ the average momentum of inertia ($\alpha=1,2$, and $N$ the total number of particles), 
$\mathbf{D}_{12}=
\langle\varepsilon_2|(\sum_{j=1}^Nm_j\,\hat{\mathbf{q}}_j)|\varepsilon_1\rangle$,
$M=\sum_{j=1}^N m_j$ the total mass, and $\Lambda$ is the coupling constant with the collapse or gravitational field. For the CSL model we get
$\Lambda_{\text{\tiny CSL}}\simeq1.12\times10^{-9}\,$s$^{-1}$ and for the DP model $\Lambda_{\text{\tiny DP}}\simeq 7.39\times10^{-25}\,$Hz~\cite{supp}. From now on, we consider only CSL as the effect of DP model is by far smaller.
\begin{table*}[t]
\begin{tabular}[b]{p{9cm} p{4cm} p{3cm}}
\hline
{\bf System} & $\beta_{\text{\tiny N}}$ (Hz) & $\gamma_{\text{\tiny N}}$ (Hz)
\\\hline
{\bf Hydrogen-like Atoms} & $10^{-20}-10^{-18}$ & $\sim10^{-53}$
\\\hline
{\bf Harmonic oscillator} & $\frac{3\Lambda}{4}\left(\frac{\mu\, x_0}{m_0\,r_C}\right)^2$ & $\frac{\Lambda^2}{32\omega_0}\left(\frac{\mu\, x_0}{m_0\,r_C}\right)^4$ 
\\
$\mu=1\,$amu and $\omega_0=10^{10}$Hz & $5.3\times10^{-13}$ & $6.2\times10^{-36}$
\\
$\mu=10^7\,$amu and $\omega_0=1.7\times10^8$Hz & $3.1\times10^{-4}$ & $1.3\times10^{-16}$
\\\hline
{\bf Double-well} & $\frac{\Lambda}{8}\left(\frac{\mu\, q_0}{m_0\,r_C}\right)^2$ & $\frac{\Lambda^2}{128\,\omega_0}\left(\frac{\mu\, q_0}{m_0\,r_C}\right)^4$
\\
$\mu=m_e=5.5\times10^{-4}\,$amu and $q_0=1$\AA & $4.2\times10^{-23}$ & $10^{-57}-10^{-55}$
\\
$\mu=1\,$amu and $q_0=1$\AA & $1.4\times10^{-16}$ & $10^{-44}-10^{-42}$
\\
$\mu=10^7\,$amu and $q_0=1$\AA & $0.014$ & $10^{-16}-10^{-18}$
\\\hline
\end{tabular}
\caption{{\bf Collapse broadening and shift as predicted by the CSL model.} Three relevant situations have been considered: the transition from the 2P to 1S state in a Hydrogen-like atom; the transition from the first excited state to the ground state for a harmonic potential, and for a double-well potential (see supplementary material for a description of these systems). The latter case is particularly relevant to describe chiral molecules. The constant $\Lambda\simeq1.12\times10^{-9}\,$Hz measures the strength of the collapse (see main text), $m_0=1\,$amu, $r_C\simeq10^{-7}\,$m is the correlation length of the noise inducing the collapse, $x_0=\sqrt{\hbar/\mu\omega_0}$ is the zero-point width of harmonic oscillator, and $q_0$ is the separation of the minima of double-well potential. For the double-well potential, we assume the range of the molecular vibration: $\omega_0\sim10^{12}-10^{14}$Hz. We have considered only the predictions of the CSL model, since those of gravity induced models (Diosi-Penrose) are much smaller. Note, that all numbers in this table are exemplary to illustrate the magnitude of the spectral effects, and not necessarily realised in experiments yet.}
\end{table*}

%%%%%%%%%%%%%%%%%%%%%%%%%%%%%%%%%%%%%%%%%%%%%%%%%%%%%%%%%%%%%%%%%%%%%%%%%%%%%%%%%
\noindent{\it Universality of broadening.}
%%%%%%%%%%%%%%%%%%%%%%%%%%%%%%%%%%%%%%%%%%%%%%%%%%%%%%%%%%%%%%%%%%%%%%%%%%%%%%%%%
-- A crucial feature of Eqs.~(\ref{eq:2}) and~(\ref{eq:3}) is that both broadening and shift induced by collapse models are universal, in the sense that they depend only on the mass of the system (at the practical level, only on the mass of those particles, whose position changes significantly during the transition) and on the geometry of the levels, and nothing else. This has to be compared with decoherence broadening and shift, which depends also on the details of the surrounding environment: mass of the bath particle, cross section, pressure, temperature. Moreover, in case of collapse models, they roughly scale  quadratically with the mass of the system, while the mass dependence with decoherence is different. This behavior represents a specific signature, which can be used to discriminate collapse broadening from decoherence broadening (see Fig.1).

%%%%%%%%%%%%%%%%%%%%%%%%%%%%%%%%%%%%%%%%%%%%%%%%%%%%%%%%%%%%%%%%%%%%%%%%%%%%%%%%%
\noindent{\it Experimental implication}.
%%%%%%%%%%%%%%%%%%%%%%%%%%%%%%%%%%%%%%%%%%%%%%%%%%%%%%%%%%%%%%%%%%%%%%%%%%%%%%%%%
-- In general, spectroscopy can be done in any domain of the electro-magnetic spectrum and any physical system with discrete energy levels is potentially good for tests. This means in principle any degree of freedom of particles: electronic, vibration or rotation, as well as collective rearrangements of atoms in bigger structures such as conformation changes of molecules or any other many-body system such as atoms in optical lattices as well as any internal degree of freedom of a condensed matter system such as spin can be used as a probe. Spectroscopy techniques with very different degree of resolution exist in such different frequency domains. Surely, to resolve the broadening effect one needs not only to resolve the spectral line centres as typically done in precision frequency spectroscopy, one also needs to resolve the width of the spectral line with the same resolution.

\noindent {\it Resolution of the spectroscopic detection system.} To detect small deviations as those shown in Table 1, one needs ultra-high resolution spectroscopic techniques. In general to observe small spectral effects the quality factor ($Q=\omega /\beta$) of the spectroscopic system has to be as high as possible. State of the art for frequency stability of a laser (1/Q) is of the order 10$^{-16}$ in the visible spectral range~\cite{Cole2013}, which can be transferred to other frequency domains in principle by frequency combs~\cite{frequencycomb}. However, most recently the frequency stability for an atomic frequency measurement in the visible/near-infrared spectral domain has been measured to resolve instabilities even on the the order of 10$^{-18}$~\cite{Hinkel2013}. Such technical capabilities are very promising to observe spectral effects.

\noindent {\it Competing decoherence effects.} In typical experiments, there are other spectral broadening effects, which dominate the shape and width of spectral lines. The effects reduce the lifetime $\tau$ of the coherently excited state, but can be maintained by controlling limiting environmental parameters as temperature and pressure, or by experimental arrangements. Exemplary we give numbers for collisional and Doppler broadening for a vibrational mode of a generic system (mass of system $M = 10^7$ amu, $\omega_0$ = 10$^{13}$ Hz at $T=10\,$ K): Collisional broadening happens when emission is triggered by collision with other particles and an analytic expression in Hard sphere approximation is~\cite{Loudon}:  $\beta_{\text{\tiny C}}=4d^2p\sqrt{\frac{\pi}{\mu_d k_{\text{\tiny B}}T}}$, with $p$ the pressure, $d$ the closest distance of colliding particles, and $\mu_d=m\,M/(M+m)$ where $m$ is the mass of bath particles. For $p=10^{-7}$ Pa, $m=28\,$amu (nitrogen) and $d=1 \text{\AA}$, we find: $\beta_{\text{\tiny C}}\simeq$ 3 mHz. Doppler broadening is an effect of the thermal motion of an ensemble of emitters and is given by~\cite{Loudon}:
$\beta_{\text{\tiny D}}=\omega_0\sqrt{\frac{2k_{\text{\tiny B}}T\ln 2}{Mc^2}}\simeq 3568\,$ Hz. However, saturation spectroscopy can be used to avoid Doppler broadening. Therefore, both effects can be kept smaller than $\beta_N$, see Tab. 1.

%%%%%%%%%%%%%%%%%%%%%%%%%%%%%%%%%%%%%%%%%%%%%%%%%%%%%%%%%%%%%%%%%%%%%%%%%%%%%%%%%%%%%5
\noindent{\it Search for the ultimate test system.}
%%%%%%%%%%%%%%%%%%%%%%%%%%%%%%%%%%%%%%%%%%%%%%%%%%%%%%%%%%%%%%%%%%%%%%%%%%%%%%%%%%%%%5
-- Non-linear spectral effects are small, but within reach of experimental observation. Thermal and collisional effects can be reduced to the required limitations. While we cannot predict the ultimate two-level system here, we give explicit examples to illustrate the relation to state-of-the-art experiments. 

Today ultra-high resolution spectroscopy in the mid-infrared spectral range is done with a precision of $10^{-13}$~\cite{Daussy1999} and planned to be improved to $10^{-16}$~\cite{Darquie2010}. We estimate spectral effects of non-linearities of the order of  $10^{-14}$ to $10^{-16}$ for a double-well system of $10^7$ amu and $q_0$= 1 \AA (see Table 1) probed in the mid-infrared range. This test seems feasible in the near future. 

Further electron or nuclear spins are known to be accessible with ultra-high frequency resolution of 10 mHz. The Lamor frequency of the system $^{7}$Li$^{+}$ FID in water, would need to be detected with a spectral resolution of about 30 $\mu$Hz to show a nonlinear effect, which is only three orders of magnitude away from todays resolution~\cite{Appelt2006}. Further relevant are solid state systems, like semiconductor microcavities, quantum dots, or nanodiamonds with vacancy centers and spin structure~\cite{Yamamoto1999} as well as opto-mechanical systems. 

\noindent{\it Conclusion.} --The mass proportional CSL model predicts a measurable spectral line broadening effect. Ultra-stable lasers as used in state-of-the-art experiments are sufficient to observe this effect, if the appropriate system can be identified, according to the significant parameters: spectral lifetime and mass. Competing line-broadening decoherence effects can be controlled by maintaining experimental conditions. In general and in contrast to decoherence effects, CSL broadening scales quadratically with mass and does not depend on experimental controllable parameters. This opens the door for systematic investigations on the hunt for intrinsic non-linearities and a possibly deeper theory to describe nature.

%%%%%%%%%%%%%%%%%%%%%%%%%%%%%%%%%%%%%%%%%%%%%%%%%%%%%%%%%%%%%%%%%%%%%%%%%%%%%%%%%%%%%5
\noindent{\it Acknowledgements.}
%%%%%%%%%%%%%%%%%%%%%%%%%%%%%%%%%%%%%%%%%%%%%%%%%%%%%%%%%%%%%%%%%%%%%%%%%%%%%%%%%%%%%5
-- MB and AB acknowledge financial support from the EU project NANOQUESTFIT. 
AB wish to thank the COST Action MP1006 ``Fundamental Problems in Quantum
Physics". AB acknowledges partial support from INFN. 
MB thanks Prof. M. Ghasem Mahjani of K. N. Toosi Univ., Iran, for his valuable supports. HU thanks the UK funding agency EPSRC (EP/J014664/1), the Foundational Questions Institute (FQXi) and the John F Templeton Foundation for financial support. 
All authors thank  Dr. B. Darqui\'e of LPL, Universit\'e Paris 13, France, for his valuable comments on experimental part of this letter.

%\newpage
\begin{widetext}

\section*{Supplementary information}

Here we provide an extensive analysis on how modifying Schr\"{o}dinger equation with non-linearities, changes the spectral density of emitted radiation from a general two-level system. We will show that the non-linearity manifests as a broadening and shift in the spectral density of emitted radiation. We first discuss how the non-linearity can be introduced into the dynamics. Then, following standard QED formulations, we obtain the broadening and the shift induced by non-linearity. We finally compute analytically and quantify these radiative corrections from two most studied collapse models.

% \noindent {\bf Modified Schr\"{o}dinger equations.}

\section*{S1: Dynamical equations.} It was first proven by Gisin~\cite{signal0} that, in order to avoid superluminal signalling, nonlinear terms can be added to the Schr\"odinger equation only combined with stochastic terms, in such a way that the equivalence relation among statistical ensembles of states is preserved by the dynamics~\cite{signal1,signal2,signal3}. In more mathematical terms, this means that the modified dynamics for the wave function must generate a closed linear dynamics for the density matrix. Given these premises, it was recently proven~\cite{signal4} that such a dynamics must be of the Lindblad type. It is important to notice that in the proof, complete positivity---usually requested in order to derive Lindblad's theorem---is not a necessary hypothesis, but comes about from the existence of a (Markovian) dynamics for the wave function.
Therefore, we start from a dynamics for the density matrix of the form:
\begin{equation}\label{Lindblad}
\frac{\text d\hat{\rho}_t}{\text dt}  =  -\frac{i}{\hbar}[\hat{H}_0,\hat{\rho}_t] + \lambda\sum_{k=1}^n\left(\hat{L}_k\hat{\rho}_t\hat{L}_k^{\dagger} - \frac12 \hat{L}_k^{\dagger}\hat{L}_k\hat{\rho}_t - \frac12 \hat{\rho}_t\hat{L}_k^{\dagger}\hat{L}_k\right)\,
\end{equation}
where the Lindblad operators $\hat{L}_k$ can describe decoherence effect or, as it is the case here, intrinsic non-linearities in the dynamics for the wave function. Collapse models induce a dynamics of this type, but here we want to stay more general. The most convenient unraveling of Eq.~\eqref{Lindblad} for solving the equations of motions, is given in terms of a stochastic potential added to the Schr\"odinger equation~\cite{stoch1,stoch2,stoch3,stoch4,stoch5}:
\begin{equation}
i\hbar \frac{d}{dt} \psi_t = \left[ \hat{H}_0 + \hat{V}_t \right ] \psi_t, \qquad \hat{V}_t = - \hbar\,\sqrt{\lambda}\sum_{k=1}^n \hat{L}_k \xi_t^{(k)}
\end{equation}
where $\xi_t^{(k)}$ are $n$ independent white noises. Here, we have assumed that the Lindblad operators $\hat{L}_k$ are self-adjoint, which is the case for most proposals for nonlinear and stochastic modifications of the Schr\"odinger equation. 

Since with violations of the superposition principle we mean superpositions in space, the Lindblad operators are taken as functions of space variables, therefore we have:
\begin{equation} \label{eq:v}
\hat{V}_t = - \hbar\,\sqrt{\lambda} \int d^3x\; \hat{L}({\bf x}) \xi_t({\bf x}),
\end{equation} 
where $\xi_t({\bf x})$ is a noise-field, white both in space and time. Note that, in this form, the dynamical equation is still linear. As discussed several times in the literature~\cite{stoch1,stoch2,stoch3,stoch4,stoch5}, the effects of nonlinear terms introduced in the Schr\"odinger equation, at the statistical level, can be mimicked also by linear random potentials. For individual realizations of the noise, the affects are very different (those of a linear dynamics vs those of a nonlinear one), while at the statistical level they coincide, if the potential is suitably chosen.
% The non-linear modifications of the Schr\"{o}dinger equation should be done in a stochastic fashion, otherwise the non-linearities allow the faster-then-light signalling~\cite{signal0,signal1,signal2,signal3}. The stochastic modifications of the Schr\"{o}dinger equation is usually called as {\it collapse models}. We will use the dynamics of collapse models where a stochastic field (usually called as {\it collapse filed}) couples non-linearly with massive systems, thus inducing the intrinsic non-linearities.
% We will formulate the spectral density of spontaneously emitted light from a two-level system that interacts with quantized vacuum field and also with fundamental collapse field of collapse models. We will see that the collapse field acts a noise source, inducing the level shift and also level widths (broadening). 
\vspace{2 mm}

\section*{S2: Two level systems.} We consider the situation in which the system's dynamics effectively involves only two levels, whose transition dipole matrix element is not zero ($\mathbf{d}_{12}=\langle\varepsilon_1|\hat{\mathbf{d}}|\varepsilon_2\rangle\neq0$, with the dipole operator defined as  $\hat{{\bf d}} = \sum_i e_i \hat{{\bf q}}_i$). This is means that the higher energy level eventually decays to the lower one, by emitting a photon. Therefore, the standard quantum Hamiltonian characterizing the interaction between a two-level system and a quantized radiation field can be, in the dipole approximation, written in the form~\cite{agar,QO_MW,QO_GC,QT_rad}:
\begin{eqnarray}
\hat{H}_0&=&\hbar
\left(
\sum_{s,\mathbf{k}} 
\omega\,\hat{a}^\dagger_{s,\mathbf{k}}\hat{a}_{s,\mathbf{k}}
+
(\omega_{0}/2)\hat{\sigma}_z
-i\omega_{0}  \sum_{s,\mathbf{k}} \left\lbrace
g_{s,\mathbf{k}}\left(\hat{\sigma}_+-\hat{\sigma}_-\right)\hat{a}_{s,\mathbf{k}} 
- \text{H.C.}\right\rbrace\right),
\end{eqnarray}
with $g_{s,\mathbf{k}} = (2\epsilon_0\hbar\omega L^3)^{-1/2}\,
\mathbf{d}_{12}\cdot\mathbf{e}_{s,\mathbf{k}}$ the coupling constant of radiation-matter, 
% $\mathbf{d}_{12}=\langle\varepsilon_1|\hat{\mathbf{d}}|\varepsilon_2\rangle$ the dipole moment element, 
$\hat{\sigma}_+ = |\varepsilon_2\rangle\langle\varepsilon_1|$, $\hat{\sigma}_- = |\varepsilon_1\rangle\langle\varepsilon_2|$, 
$\hat{\sigma}_z=\hat{\sigma}_+\hat{\sigma}_--\hat{\sigma}_-\hat{\sigma}_+$, $\omega_0=(\varepsilon_2-\varepsilon_1)/\hbar$
and $\{|\varepsilon_1\rangle,\,|\varepsilon_2\rangle\}$ the two levels of matter. 
All other terms have the usual meanings. 

The two-level representation of $\hat{V}_t$ is obtained by calculating $\langle\varepsilon_\alpha|\hat{V}_t|\varepsilon_\beta\rangle$ with $\alpha,\beta=1,2$. 
In general, one has:
\begin{equation}
\hat{V}_t = -\hbar\left(
\sqrt{\lambda_z}\,w_t^{(z)}\,\hat{\sigma}_z
+\sqrt{\lambda_x}\,w_t^{(x)}\,\hat{\sigma}_x+\sqrt{\lambda_x}\,w_t^{(y)}\,\hat{\sigma}_y
\right)
\end{equation}
where $\lambda_i$ ($i=x,y,z$) are collapse rates and $w^{(i)}_t$ are three white noises. Eigenenergies are real functions in most cases, thus one finds: $\lambda_y=0$. Therefore $\hat{V}_t$ simplifies to: 
\begin{eqnarray}
\label{eq:noise}
\hat{V}_t&=&- \hbar\left(
\sqrt{\lambda_z}\,w_t^{(z)}\,\hat{\sigma}_z+\sqrt{\lambda_x}\,w_t^{(x)}\,\hat{\sigma}_x
\right).
\end{eqnarray}

\section*{S3: Solution of the equations of motion: shift and broadening.} 
The radiative corrections of the non-linearities appear in a very natural way from the formulation of the spectral density of emitted light~\cite{agar,QO_MW,QO_GC,QT_rad} which is given by the stochastic expectation (averaging) of the Fourier transform of the normalized dipole-dipole autocorrelation function. 
The dipole-dipole autocorrelation function is given by $\langle\hat{\sigma}_+(t+\tau)\hat{\sigma}_-(t)\rangle$.
Accordingly, in order to obtain the spectral density, we will pursue the following steps. First, we will solve the Heisenberg equations of systems operator. Then, we will obtain the corresponding differential equations for dipole-dipole autocorrelation functions. We will solve these equations  
using our previous results. With new solutions, we will finally get the dipole-dipole autocorrelation functions, and thus the spectral density.

With the Hamiltonian $\hat{H}=\hat{H}_0+\hat{V}_t$, the Heisenberg equations of motion for the system operators take the forms:
\begin{eqnarray}
\label{eq:sigma_z}
\frac{d}{dt}\hat{\sigma}_z(t)&=&
-2\beta_{\text{\tiny QED}}\left(\hat{\sigma}_z(t)+\hat{I}\right)
-2\omega_0 \left(\hat{\sigma}_x(t)\,
\mathbf{d}_{12}\cdot\hat{\mathbf{A}}^{(+)}_{\text{\tiny free}}(0,t)
+\mathbf{d}_{12}\cdot\hat{\mathbf{A}}^{(-)}_{\text{\tiny free}}(0,t)\,\hat{\sigma}_x(t)\right)
-2\sqrt{\lambda_x}\,w^{(x)}_t\,\hat{\sigma}_y(t),
\\
\label{eq:sigma_-}
\frac{d}{dt}\hat{\sigma}_y(t)&=&
-\left(\Omega_{\text{\tiny QED}}-2\sqrt{\lambda_z}\,w^{(z)}_t\right)\hat{\sigma}_x(t)
-\beta_{\text{\tiny QED}}\hat{\sigma}_y(t)
+2\sqrt{\lambda_x}\,w^{(x)}_t\,\hat{\sigma}_z(t),
\\
\label{eq:sigma_+}
\frac{d}{dt}\hat{\sigma}_x(t)&=&
-\left(\Omega_{\text{\tiny QED}}-2\sqrt{\lambda_z}\,w^{(z)}_t\right)\hat{\sigma}_y(t)
-\beta_{\text{\tiny QED}}\,\hat{\sigma}_x(t)
+2\omega_0
\left(\hat{\sigma}_z(t)\,
\mathbf{d}_{12}\cdot\hat{\mathbf{A}}^{(+)}_{\text{\tiny free}}(0,t)
+\mathbf{d}_{12}\cdot\hat{\mathbf{A}}^{(-)}_{\text{\tiny free}}(0,t)\,\hat{\sigma}_z(t)
\right),
\end{eqnarray}
where last terms on the right hand side of Eqs.~\eqref{eq:sigma_z} and \eqref{eq:sigma_-} are the new contributions to the dynamics due to the noise terms describing nonlinear effects. Other terms are obtained using standard derivations given by quantum statistical theories of spontaneous emission~\cite{agar,QO_MW,QO_GC,QT_rad}. In the above equations, $\Omega_{\text{\tiny QED}} = \omega_0-\gamma_{\text{\tiny QED}}$ where
\begin{eqnarray}
\gamma_{\text{\tiny QED}}&=&
\frac{\omega_0^2|\mathbf{d}_{12}|^2}{3\epsilon_0\hbar\pi^2c^3}
\left(\mathcal{P}\int_0^\infty \frac{d\omega \,\omega}{\omega-\omega_0}
-\mathcal{P}\int_0^\infty \frac{d\omega \,\omega}{\omega+\omega_0}\right),
\end{eqnarray} 
is the Lamb shift, with $\mathcal{P}$ is the Cauchy principal part, which can be normalized in the standard fashion~\cite{bethe}; the parameter
\begin{eqnarray}
\beta_{\text{\tiny QED}}&=&\frac{\omega_0^3|\mathbf{d}_{12}|^2}{6\pi\epsilon_0\hbar c^3}
\end{eqnarray}
is the standard spontaneous emission rate; and:
\begin{eqnarray}
\mathbf{d}_{12}\cdot\hat{\mathbf{A}}^{(+)}_{\text{\tiny free}}(0,t)&=&
\sum_{s,\mathbf{k}} g_{s,\mathbf{k}}\,e^{-i\omega t} \,
\hat{a}_{s,\mathbf{k}}(0).
\end{eqnarray}

We now average Eqs.~(\ref{eq:sigma_z}) and~(\ref{eq:sigma_-}) over the initial state $|\psi\rangle|0\rangle$, according to which matter is in a generic state $|\psi\rangle$ and the radiation field in the vacuum state. 
Therefore we get:
\begin{eqnarray}
\frac{d}{dt}\langle\hat{\sigma}_z(t)\rangle&=&
-2\beta_{\text{\tiny QED}}\left(\langle\hat{\sigma}_z(t)\rangle+1\right)
-2\sqrt{\lambda_x}\,w^{(x)}_t\,\langle\hat{\sigma}_y(t)\rangle
\\
\frac{d}{dt}\langle\hat{\sigma}_y(t)\rangle&=&
\left(\Omega_{\text{\tiny QED}}-2\sqrt{\lambda_z}\,w^{(z)}_t\right)\langle\hat{\sigma}_x(t)\rangle
-\beta_{\text{\tiny QED}}\langle\hat{\sigma}_y(t)\rangle
+2\sqrt{\lambda_x}\,w^{(x)}_t\,\langle\hat{\sigma}_z(t)\rangle
\\
\frac{d}{dt}\langle\hat{\sigma}_x(t)\rangle&=&
-\left(\Omega_{\text{\tiny QED}}-2\sqrt{\lambda_z}\,w^{(z)}_t\right)\langle\hat{\sigma}_y(t)\rangle
-\beta_{\text{\tiny QED}}\langle\hat{\sigma}_x(t)\rangle.
\end{eqnarray}
The above stochastic differential equations should be understood in Stratonovich sense. Since we want to compute stochastic averages, it is more convenient to switch to the It\^o formalism. To this end, one can use Eqs.~(10.2.5) to~(10.2.7) of Ref.~\cite{sto_book}. Then, once expressed in the It\^{o} form, by using theorem (8.5.5) of Ref.~\cite{sto_book}, one can prove that the stochastic expectations satisfy the following equations:
\begin{eqnarray}
\frac{d}{dt}\mathbb{E}(\langle\hat{\sigma}_z(t)\rangle)&=&
-2(\beta_{\text{\tiny QED}}+\lambda_x)\,\mathbb{E}(\langle\hat{\sigma}_z(t)\rangle)-2\beta_{\text{\tiny QED}}
\\
\frac{d}{dt}\mathbb{E}(\langle\hat{\sigma}_y(t)\rangle)&=&
\Omega_{\text{\tiny QED}}\,\mathbb{E}(\langle\hat{\sigma}_x(t)\rangle)
-(\beta_{\text{\tiny QED}}+2\lambda_x+2\lambda_z)\,\mathbb{E}(\langle\hat{\sigma}_y(t)\rangle)
\\
\frac{d}{dt}\mathbb{E}(\langle\hat{\sigma}_x(t)\rangle)&=&
-\Omega_{\text{\tiny QED}}\,\mathbb{E}(\langle\hat{\sigma}_y(t)\rangle)
-(\beta_{\text{\tiny QED}}+2\lambda_z)\mathbb{E}(\langle\hat{\sigma}_x(t)\rangle)
\end{eqnarray}
Using the solutions of above equations, one can find:
\begin{eqnarray}
\label{eq:decay}
\mathbb{E}(\langle\hat{\sigma}_z(t)\rangle)&=&
\left(
\frac{\beta_{\text{\tiny QED}}}{\beta_{\text{\tiny QED}}+\lambda_x}+\langle\hat{\sigma}_z(0)\rangle
\right)e^{-2(\beta_{\text{\tiny QED}}+\lambda_x)t}
-\frac{\beta_{\text{\tiny QED}}}{\beta_{\text{\tiny QED}}+\lambda_x},
%\qquad\text{\small Change in the energy of system}
\\
\label{eq:intensity0}
\mathbb{E}(\langle\hat{\sigma}_+(t)\hat{\sigma}_-(t)\rangle)&=&
\frac{1}{2}\left[\left(
\frac{\beta_{\text{\tiny QED}}}{\beta_{\text{\tiny QED}}+\lambda_x}+\langle\hat{\sigma}_z(0)\rangle
\right)e^{-2(\beta_{\text{\tiny QED}}+\lambda_x)t}
+\frac{\lambda_x}{\beta_{\text{\tiny QED}}+\lambda_x}\right],
%\qquad\text{\small Population of excited state}
\end{eqnarray}
where $\hbar\omega_0\,\mathbb{E}(\langle\hat{\sigma}_z(t)\rangle)$ gives the rate of energy emission by matter; and $\mathbb{E}(\langle\hat{\sigma}_+(t)\hat{\sigma}_-(t)\rangle)$ represents the change in the population of the excited state $|\varepsilon_2\rangle$. When the initial state is $|\varepsilon_1\rangle$, we have: $\langle\hat{\sigma}_z(0)\rangle=-1$, and for $|\varepsilon_2\rangle$, we have:
$\langle\hat{\sigma}_z(0)\rangle=1$. On the other hand, for both initial states $|\varepsilon_{1,2}\rangle$ we get: 
\begin{eqnarray}
\label{eq:in-val}
\mathbb{E}(\langle\hat{\sigma}_-(t)\rangle)=\mathbb{E}(\langle\hat{\sigma}_+(t)\rangle)=0&\text{for}&|\varepsilon_{1,2}\rangle|0\rangle.
\end{eqnarray}

Using Eq.~\eqref{eq:intensity0}, one can compute for example the mean light intensity of emitted radiation in the far-field limit~\cite{agar,QO_MW,QO_GC,QT_rad} as follows:
\begin{eqnarray}\label{eq:intensity}
\langle\hat{I}(\mathbf{r},t)\rangle &=&
\left(\frac{\omega_0^2|\mathbf{d}_{12}|}{8\pi\varepsilon_0c^2r}\right)^2
\left(1-\frac{1}{2}\sin^2\theta\right)
\left[\left(\frac{\beta_{\text{\tiny QED}}}{\beta_{\text{\tiny QED}}+\lambda_x}+\langle\hat{\sigma}_z(0)\rangle
\right)e^{-2(\beta_{\text{\tiny QED}}+\lambda_x)(t-\frac{r}{c})}
+\frac{\lambda_x}{\beta_{\text{\tiny QED}}+\lambda_x}\right],\quad t>\frac{r}{c},
\end{eqnarray}
with $\theta$ the polar angle of the $\mathbf{r}$-vector, and the complex dipole moment $\mathbf{d}_{12}$ lies in the $xy$-plane, where $\mathbf{r}$ is the vector connecting the center-of-mass of system to the detector. $\langle\hat{I}(\mathbf{r},t)\rangle$ is a very interesting quantity for experimental research; however, here we are concerned with the spectral density of emitted light, which we now compute.

The exponential nature of energy decay, as given by Eqs.~(\ref{eq:decay}) and~(\ref{eq:intensity}), suggests that the spectral distribution of the emitted radiation is Lorentzian. The explicit mathematical form of spectral density can be obtained by computing the dipole-dipole autocorrelation function, $\mathbb{E}(\langle\hat{\sigma}_+(t+\tau)\hat{\sigma}_-(t)\rangle)$, and then using the Wiener-Khinchin theorem~\cite{QT_rad}. The time derivative of this autocorrelation function can be obtained by making the change $t\rightarrow t+\tau$ in Eqs.~(\ref{eq:sigma_z}), (\ref{eq:sigma_-}) and~(\ref{eq:sigma_+}), and then writing down the derivatives with respect to $\tau$. After multiplying the result from the right with $\hat{\sigma}_-(t)$ and then taking the quantum average over the initial state $|\psi\rangle|0\rangle$, we find:
%\begin{eqnarray}
%\frac{d}{d\tau}\langle\hat{\sigma}_z(t+\tau)\hat{\sigma}_-(t)\rangle&=&
%-2\beta_{\text{\tiny QED}}\left(\langle\hat{\sigma}_z(t+\tau)\hat{\sigma}_-(t)\rangle
%+\langle\hat{\sigma}_-(t)\rangle\right)
%-2\sqrt{\lambda}\,w^{(x)}_\tau\,\langle\hat{\sigma}_y(t+\tau)\hat{\sigma}_-(t)\rangle,
%\\
%\frac{d}{d\tau}\langle\hat{\sigma}_-(t+\tau)\hat{\sigma}_-(t)\rangle&=&
%-\left(i\left(\Omega_{\text{\tiny QED}}-2\sqrt{\lambda_z}\,w^{(z)}_\tau\right)+\beta_{\text{\tiny QED}}\right)
%\langle\hat{\sigma}_-(t+\tau)\hat{\sigma}_-(t)\rangle
%-i\sqrt{\lambda_x}\,w^{(x)}_\tau\,
%\langle\hat{\sigma}_z(t+\tau)\hat{\sigma}_-(t)\rangle,
%\\
%\frac{d}{d\tau}\langle\hat{\sigma}_+(t+\tau)\hat{\sigma}_-(t)\rangle&=&
%\left(i\left(\Omega_{\text{\tiny QED}}-2\sqrt{\lambda_z}\,w^{(z)}_\tau\right)-\beta_{\text{\tiny QED}}\right)
%\langle\hat{\sigma}_+(t+\tau)\hat{\sigma}_-(t)\rangle
%+i\sqrt{\lambda}\,w^{(x)}_\tau\,
%\langle\hat{\sigma}_z(t+\tau)\hat{\sigma}_-(t)\rangle.
%\end{eqnarray}
%Like before, we switch to the corresponding equations for the Pauli matrices:
\begin{eqnarray}
\frac{d}{d\tau}\langle\hat{\sigma}_z(t+\tau)\hat{\sigma}_-(t)\rangle&=&
-2\beta_{\text{\tiny QED}}\left(\langle\hat{\sigma}_z(t+\tau)\hat{\sigma}_-(t)\rangle
+\langle\hat{\sigma}_-(t)\rangle\right)
-2\sqrt{\lambda_x}\,w^{(x)}_\tau\,\langle\hat{\sigma}_y(t+\tau)\hat{\sigma}_-(t)\rangle,
\\
\frac{d}{d\tau}\langle\hat{\sigma}_y(t+\tau)\hat{\sigma}_-(t)\rangle&=&
\left(\Omega_{\text{\tiny QED}}-2\sqrt{\lambda_z}\,w^{(z)}_\tau\right)\langle\hat{\sigma}_x(t+\tau)\hat{\sigma}_-(t)\rangle
-\beta_{\text{\tiny QED}}\langle\hat{\sigma}_y(t+\tau)\hat{\sigma}_-(t)\rangle\\\nonumber&&
+2\sqrt{\lambda_x}\,w^{(x)}_\tau\,
\langle\hat{\sigma}_z(t+\tau)\hat{\sigma}_-(t)\rangle,
\\
\frac{d}{d\tau}\langle\hat{\sigma}_x(t+\tau)\hat{\sigma}_-(t)\rangle&=&
-\beta_{\text{\tiny QED}}\langle\hat{\sigma}_x(t+\tau)\hat{\sigma}_-(t)\rangle
-\left(\Omega_{\text{\tiny QED}}-2\sqrt{\lambda_z}\,w^{(z)}_t\right)
\langle\hat{\sigma}_y(t+\tau)\hat{\sigma}_-(t)\rangle.
\end{eqnarray}
Using the aforementioned theorems to switch between Stratonovich and It\^{o} forms and also to obtain the stochastic expectations, and also taking into the account the result given by Eq.\eqref{eq:in-val}, we find:
\begin{eqnarray}
\frac{d}{d\tau}\mathbb{E}(\langle\hat{\sigma}_z(t+\tau)\hat{\sigma}_-(t)\rangle)&=&
-2(\beta_{\text{\tiny QED}}+\lambda_x)\,\mathbb{E}(\langle\hat{\sigma}_z(t+\tau)\hat{\sigma}_-(t)\rangle)
\\
\frac{d}{d\tau}\mathbb{E}(\langle\hat{\sigma}_y(t+\tau)\hat{\sigma}_-(t)\rangle)&=&
\Omega_{\text{\tiny QED}}\,\mathbb{E}(\langle\hat{\sigma}_x(t+\tau)\hat{\sigma}_-(t)\rangle)
-(\beta_{\text{\tiny QED}}+2\lambda_x+2\lambda_z)\,\mathbb{E}(\langle\hat{\sigma}_y(t+\tau)\hat{\sigma}_-(t)\rangle)
\\
\frac{d}{d\tau}\mathbb{E}(\langle\hat{\sigma}_x(t+\tau)\hat{\sigma}_-(t)\rangle)&=&
-(\beta_{\text{\tiny QED}}+2\lambda_z)\,\mathbb{E}(\langle\hat{\sigma}_x(t+\tau)\hat{\sigma}_-(t)\rangle)
-\Omega_{\text{\tiny QED}}\,\mathbb{E}(\langle\hat{\sigma}_y(t+\tau)\hat{\sigma}_-(t)\rangle)
\end{eqnarray}
Accordingly, for the dipole-dipole autocorrelation function we get:
\begin{eqnarray}
\mathbb{E}(\langle\hat{\sigma}_+(t+\tau)\hat{\sigma}_-(t)\rangle)
&=&
e^{-(\beta_{\text{\tiny QED}}+\lambda_x+2\lambda_z)\tau}
\left(
\cosh(i\Omega\tau)+\frac{\Omega_{\text{\tiny QED}}\sinh(i\Omega\tau)}{\Omega}
\right)
\mathbb{E}(\langle\hat{\sigma}_+(t)\hat{\sigma}_-(t)\rangle)
\\
&=&
e^{-(\beta_{\text{\tiny QED}}+\lambda_x+2\lambda_z)\tau}
\left(
\cos\Omega \tau+i\frac{\Omega_{\text{\tiny QED}}}{\Omega}\sin \Omega\tau
\right)
\mathbb{E}(\langle\hat{\sigma}_+(t)\hat{\sigma}_-(t)\rangle)
\end{eqnarray}
with $\Omega=\sqrt{{\Omega_{\text{\tiny QED}}}^2-\lambda_x^2}$ for $\Omega_{\text{\tiny QED}}>\lambda_x$. For $\Omega_{\text{\tiny QED}}\gg\lambda_x$, which is the case for most cases of experimental interest, we can expand $\Omega_{\text{\tiny QED}}/\Omega$ to the first leading term in $\lambda_x/\Omega_{\text{\tiny QED}}$, and we get:
\begin{eqnarray}
\mathbb{E}(\langle\hat{\sigma}_+(t+\tau)\hat{\sigma}_-(t)\rangle)
&\simeq&
e^{-(\beta_{\text{\tiny QED}}+\lambda_x+2\lambda_z)\tau}\,e^{i\Omega\tau}\,
\mathbb{E}(\langle\hat{\sigma}_+(t)\hat{\sigma}_-(t)\rangle)
\end{eqnarray}
Using Wiener-Khinchin relation, for the spectral density of emitted radiation we obtain:
\begin{eqnarray}
\mathcal{S}(\omega)&=&
\frac{1}{\pi}\left(
\frac{\beta_{\text{\tiny QED}}+\lambda_x+2\lambda_z}{(\beta_{\text{\tiny QED}}+\lambda_x+2\lambda_z)^2+(\omega-\Omega)^2}
\right) \; \equiv \; \frac{1}{\pi}\left(
\frac{\beta}{\beta^2+(\omega-\Omega)^2}\right)\,,
\end{eqnarray}
with the full width at half-maximum of $\beta$ where 
\begin{equation}
\beta = \beta_{\text{\tiny QED}} + \beta_{\text{\tiny N}}, \quad\text{with}\quad \beta_{\text{\tiny N}} = \lambda_x+2\lambda_z
\end{equation}
and the central frequency of 
\begin{eqnarray}
\Omega=\sqrt{{\Omega_{\text{\tiny QED}}}^2-\lambda_x^2}\simeq \Omega_{\text{\tiny QED}}-\frac{\lambda_x^2}{2\Omega_{\text{\tiny QED}}}\simeq \Omega_{\text{\tiny QED}}-\frac{\lambda_x^2}{2\omega_0}
=\Omega_{\text{\tiny QED}}-\Omega_{\text{\tiny N}},\quad \text{with}\quad
\Omega_{\text{\tiny N}}=\frac{\lambda_x^2}{2\omega_0}.
\end{eqnarray} 
Accordingly, the radiative corrections given by violations of the quantum superposition principle produce two observable effects: a frequency-shift and a line broadening, whose magnitude is controlled by the rates $\lambda_{x,z}$. Their numerical value depends on the specific model used to describe nonlinear (collapse) effects. 
\vspace{2 mm}

\section*{S4: Calculation of  rates $\lambda_{x,z}$}.\\
We now derive the rates $\lambda_{x,z}$ as predicted by the two most-studied collapse models in the literature: the mass proportional Continuous Spontaneous Localization (CSL) model~\cite{Cslmass}, and the Di\'{o}si-Penrose gravitational (DP) model~\cite{d,p}. 

\noindent {\it Rates for the CSL model}.
The stochastic potential $\hat{V}_t$ associated to the CSL model is~\cite{Grw,Csl,Cslmass}:
\begin{eqnarray}
\label{eq:noise-csl}
\hat{V}_t&=&-\frac{\hbar\,\sqrt{\gamma}}{m_0}\int\,d\mathbf{x}\,\xi_t(\mathbf{x})\hat{L}(\mathbf{x}), \quad \text{with}\quad \hat{L}({\bf x})=
\int d\mathbf{y}\,g\left(\mathbf{x-y}\right)\,
\sum_{\mathfrak{j}}\,m_\mathfrak{j}\,\sum_s\,\hat{a}_{\mathfrak{j}}^{\dagger}(s,\mathbf{y})\hat{a}_{\mathfrak{j}}\left(s,\mathbf{y}\right),
\end{eqnarray} 
where $m_0=1\,$amu, $\gamma \simeq 10^{-22}\text{cm}^{3}\text{s}^{-1}$~\cite{adlerphoto}, 
$\xi_t(\mathbf{x})$ is a white noise with correlation $\mathbb{E}(\xi_t(\mathbf{x})\xi_{\tau}(\mathbf{y}))=\delta(t-\tau)\,\delta(\mathbf{x}-\mathbf{y})$,
and
$\hat{a}_{\mathfrak{j}}\left(s,\mathbf{y}\right)$ is the annihilation operator of the particle type-$\mathfrak{j}$ with mass $m_\mathfrak{j}$ and the spin $s$ at position $\mathbf{y}$; and
$g({\bf r}) = \exp(-\mathbf{r}^{2}/2r_{C}^{2})/(\sqrt{2\pi}r_{C})^{3}$
with $r_C \simeq 10^{-5}\text{cm}$ the correlation length. %As we saw, the model contains two parameters: $\gamma$ and $r_C$. The first parameter measures the strength of the collapse process for a single nucleon. According to Adler~[...], its numerical value is: $\gamma \simeq ... $. Previously, a much weaker value was chosen by Ghirardi, Pearle and Rimini~[CSL]: $\gamma \simeq ...$. In the following, we will always refer to the stronger value of Adler~[...]. The second parameter $r_C$ sets the scale above which spatial superpositions become unstable and decay towards a localized state. Its numerical value is: $r_C \simeq ...$. 

In the two-level representation,
the matrix elements of $\hat{V}_t$ are given by:
\begin{eqnarray}
\label{eq:element}
V^{\alpha\beta}_t=\langle\varepsilon_\alpha|\hat{V}_t|\varepsilon_\beta\rangle
&=&
-\frac{\hbar\,\sqrt{\gamma}}{m_0}\,\int\, d\mathbf{Q}%d\mathbf{q}_1\cdots d\mathbf{q}_N\,
\,\psi_\alpha(\mathbf{Q})%\mathbf{q}_1,\cdots,\mathbf{q}_N)
\,\psi_\beta(\mathbf{Q})%\mathbf{q}_1,\cdots,\mathbf{q}_N)
\,\int\,d\mathbf{x}\,\xi_t(\mathbf{x})\,\sum_{j=1}^N\,m_j\,g(\mathbf{x}-\mathbf{q}_j),
\end{eqnarray}
with $\alpha,\beta=1,2$, and $\psi_{\alpha}(\mathbf{Q})=\langle\mathbf{Q}|\varepsilon_{\alpha}\rangle$ where we use improper states:
$|\mathbf{Q}\rangle\equiv
|\mathbf{q}_{1};\mathbf{q}_{2};\cdots;\mathbf{q}_{N}\rangle$ (with $\mathbf{q}_{j}$ the position of $j$-th particle) for which we have:
$\hat{L}(\mathbf{x})|\mathbf{Q}\rangle=
\left(\sum_{j=1}^{N}m_j\,g(\mathbf{x}-\mathbf{q}_{j})\right)|\mathbf{Q}\rangle$.
We also assume that the wave functions $\psi_{\alpha}$ are real.
Since the right side of Eq.~\eqref{eq:element} contains a Gaussian white noise, 
% of the form: $-\sqrt{\lambda}\,w_t$ with $\mathbb{E}(w_t)=0$ and $\mathbb{E}(w_{t_1}w_{t_2})=\delta(t_1-t_2)$. Accordingly, 
$\lambda_{x,z}$ can be calculated as follow:
\begin{eqnarray}
\label{eq:lambda}
\mathbb{E}\left(V^{\alpha\beta}_{t_1}V^{\alpha'\beta'}_{t_2}\right)&=&
\frac{\delta(t_1-t_2)\,\hbar^2\,\gamma}{8\pi^{3/2}r_C^3}\int d\mathbf{Q}\,d\mathbf{Q}'\,
\psi_\alpha(\mathbf{Q})\,\psi_\beta(\mathbf{Q})\,
\psi_{\alpha'}(\mathbf{Q}')\,\psi_{\beta'}(\mathbf{Q}')\,
\sum_{j,l=1}^N\frac{m_j\,m_l}{m_0^2}\,\exp[-\frac{(\mathbf{q}_j-\mathbf{q}'_l)^2}{4r_C^2}],
\\\nonumber&=&\delta(t_1-t_2)\,\hbar^2\,\lambda^{\alpha\beta}_{\alpha'\beta'}.
\end{eqnarray}
We consider the situation where the effective size of the region in which $\psi_{1,2}$ is different from zero is smaller than $r_C\simeq10^{-7}$m, which is the case for atomic and molecules systems. This is the small scale limit of the CSL model. Accordingly, by expanding the exponential term in Eq.~\eqref{eq:lambda} to  first order in $(\mathbf{q}_j-\mathbf{q}'_l)^2/4r_C^2$ and then by performing the integrations, we get:
\begin{eqnarray}
\label{eq:lambda_ap}
\lambda_{11}^{11}&\simeq&\frac{\Lambda_{\text{\tiny CSL}}\,M^2}{m_0^2}
\left(
1-
\frac{1}{M}\,\int d\mathbf{Q}\,|\psi_1(\mathbf{Q})|^2\,
\sum_{j}m_j\left(\frac{\mathbf{q}_j}{2r_C}\right)^2\right),
\\
\lambda_{22}^{22}&\simeq&\frac{\Lambda_{\text{\tiny CSL}}\,M^2}{m_0^2}
\left(
1-
\frac{1}{M}\,\int d\mathbf{Q}\,|\psi_2(\mathbf{Q})|^2\,
\sum_{j}m_j\left(\frac{\mathbf{q}_j}{2r_C}\right)^2\right),
\\
\lambda_{12}^{12}=\lambda_{21}^{21}&\simeq&\frac{\Lambda_{\text{\tiny CSL}}}{2r_C^2\,m_0^2}
\left(
\int d\mathbf{Q}\,\psi_1(\mathbf{Q})\,\psi_2(\mathbf{Q})
\sum_{j}m_j \mathbf{q}_j
\right)^2,
%\\\nonumber&=&
%\frac{\gamma}{16\pi^{3/2}r_C^5m_0^2}
%\left(\int d\mathbf{q}\,\psi^*_1(\mathbf{q})\,\psi_2(\mathbf{q})\,
%\sum_{j}\,m_j\,\mathbf{q}_j\right)^2.
\end{eqnarray}
with $\Lambda_{\text{\tiny CSL}}=\gamma/(8\pi^{3/2}r_C^3)=1.12\times10^{-9}\,$s$^{-1}$ and $M=\sum_jm_j$. In the derivation of above equations, we used the parity considerations and the orthogonality of $\psi_1$ and $\psi_2$.
Accordingly, we have: 
\begin{eqnarray}
\label{eq:l_z}
\lambda_z&=&\frac{\lambda_{11}^{11}-\lambda_{22}^{22}}{2}=
\frac{\Lambda_{\text{\tiny CSL}}\,M}{8r_C^2\,m_0^2}
\,\int d\mathbf{Q}\,\left(|\psi_2(\mathbf{Q})|^2-|\psi_1(\mathbf{Q})|^2\right)\,
\sum_{j}m_j\,\mathbf{q}_j^2,
\\
\label{eq:l_x}
\lambda_x&=&\lambda^{12}_{12}=\frac{\Lambda_{\text{\tiny CSL}}}{2r_C^2\,m_0^2}
\left(
\int d\mathbf{Q}\,\psi_1(\mathbf{Q})\,\psi_2(\mathbf{Q})
\sum_{j}m_j \mathbf{q}_j
\right)^2.
\end{eqnarray}

\section*{S5: Rates for the Di\'{o}si-Penrose (DP) model}.
The stochastic potential in the DP model is given by:
\begin{eqnarray}
\label{eq:noise-dp}
\hat{V}_t&=& - \hbar \int\,d\mathbf{x}\,\xi_t(\mathbf{x})\hat{L}(\mathbf{x}),
\end{eqnarray} 
where $\hat{L}(\mathbf{x})$ is the same like the CSL model, and $\xi_t(\mathbf{x})$ is a white noise with correlation $\mathbb{E}(\xi_t(\mathbf{x})\xi_s(\mathbf{y}))=G\,\delta(t-s)/\hbar\,|\mathbf{x}-\mathbf{y}|$ with $G$ the gravitational constant. This form of $\hat{L}(\mathbf{x})$ is different from Di\'{o}si's original proposal~\cite{d}. To avoid the divergence due to the Newton self-energy, Di\'{o}si initially proposed a Lindblad operator with a length cutoff equal to the nuclear size. Then, it was shown~\cite{ggr} that with this cutoff, predictions of the DP model are in contradiction with known observations. To avoid this problem, Ghirardi, Grassi and Rimini~\cite{ggr} proposed this new form of Lindblad operator whose cutoff is $r_C\simeq10^{-7}\,$m. This adjustment of the model was eventually acknowledged by Di\'{o}si~\cite{braz}.

For the matrix elements of $\hat{V}_t$ in two-level representation, one gets:
\begin{eqnarray}
\label{eq:element-dp}
V^{\alpha\beta}_t=\langle\varepsilon_\alpha|\hat{V}_t|\varepsilon_\beta\rangle
&=&\hbar
\,\int\, d\mathbf{Q}%d\mathbf{q}_1\cdots d\mathbf{q}_N\,
\,\psi_\alpha(\mathbf{Q})%\mathbf{q}_1,\cdots,\mathbf{q}_N)
\,\psi_\beta(\mathbf{Q})%\mathbf{q}_1,\cdots,\mathbf{q}_N)
\,\int\,d\mathbf{x}\,\xi_t(\mathbf{x})\,\sum_{j=1}^N\,m_j\,g(\mathbf{x}-\mathbf{q}_j),
\end{eqnarray}
Following the same approach that we used for the CSL model, we find:
\begin{eqnarray}
\label{eq:lambda-dp}
\mathbb{E}\left(V^{\alpha\beta}_{t_1}V^{\alpha'\beta'}_{t_2}\right)&=&
\delta(t_1-t_2)\,G\,\hbar\,\int d\mathbf{Q}\,d\mathbf{Q}'\,
\psi_\alpha(\mathbf{Q})\,\psi_\beta(\mathbf{Q})\,
\psi_{\alpha'}(\mathbf{Q}')\,\psi_{\beta'}(\mathbf{Q}')\times
\\\nonumber&&
\sum_{j,l}m_j\,m_l\,\int \frac{d\mathbf{x}\,d\mathbf{x}'}
{|\mathbf{x}-\mathbf{x}'|}\,
g(\mathbf{x}-\mathbf{q}_j)\,g(\mathbf{x}'-\mathbf{q}'_l)
\\\label{eq:erf}&=&
\frac{\delta(t_1-t_2)\,G\,\hbar}{4\pi}\int d\mathbf{Q}\,d\mathbf{Q}'\,
\psi_\alpha(\mathbf{Q})\,\psi_\beta(\mathbf{Q})\,
\psi_{\alpha'}(\mathbf{Q}')\,\psi_{\beta'}(\mathbf{Q}')\,
\sum_{j,l}m_j\,m_l\,\int d\mathbf{x}\,
g(\mathbf{x})\,\Phi_{jl}(\mathbf{x})
\\\nonumber
&=&\delta(t_1-t_2)\,\,\hbar^2\,\lambda^{\alpha\beta}_{\alpha'\beta'}
\end{eqnarray}
with
\begin{eqnarray}
\Phi_{jl}(\mathbf{x})&=&
\frac{\text{erf}\left(
\frac{|\mathbf{x}-(\mathbf{q}_j-\mathbf{q}'_l)|}{\sqrt{2}r_C}\right)}
{|\mathbf{x}-(\mathbf{q}_j-\mathbf{q}'_l)|},
\end{eqnarray}
where erf is the error function. 
Here $\Phi_{jl}(\mathbf{x})$ is slowly varying with respect to $g(\mathbf{x})$ (for more detail, see Section (4.2) of Ref.~\cite{jack}). Therefore $g(\mathbf{x})$ acts like a Dirac-delta, practically selecting the value of $\Phi_{jl}(\mathbf{x})$ in the origin $x = 0$. Accordingly, we can write:
%can be Taylor expanded around origin, that is to say~\cite{jack}:
%\begin{eqnarray}
%\Phi_{jl}(\mathbf{x})=\Phi_{jl}(0)+\mathbf{x}\cdot\nabla{\Phi}_{jl}(0)+\cdots
%\end{eqnarray} 
% Considering the first term of expansion, one gets:
\begin{eqnarray}
\label{eq:lambda-dp-app}
\mathbb{E}\left(V^{\alpha\beta}_{t_1}V^{\alpha'\beta'}_{t_2}\right)&\simeq&
\frac{\delta(t_1-t_2)\,G\,\hbar}{4\pi}
\int d\mathbf{Q}\,d\mathbf{Q}'\,
\psi_\alpha(\mathbf{Q})\,\psi_\beta(\mathbf{Q})\,
\psi_{\alpha'}(\mathbf{Q}')\,\psi_{\beta'}(\mathbf{Q}')\,
\sum_{j,l}m_j\,m_l\,\frac{\text{erf}
\left(\frac{|\mathbf{q}_j-\mathbf{q}'_l|}{\sqrt{2}r_C}\right)}
{|\mathbf{q}_j-\mathbf{q}'_l|}
\end{eqnarray}
Like before, we are interested in the cases where the spatial width of eigenenergies $\psi_{1,2}$ are smaller than $r_C$, meaning $|\mathbf{q}_j-\mathbf{q}'_l|\ll r_C$. 
This implies that $\Phi_{jl}(\mathbf{0})$ can be Taylor expanded, and to the leading order, one finds:
\begin{eqnarray}
\frac{\text{erf}
\left(\frac{|\mathbf{q}_j-\mathbf{q}'_l|}{\sqrt{2}r_C}\right)}
{|\mathbf{q}_j-\mathbf{q}'_l|}\simeq
\frac{2}{r_C\sqrt{2\pi}}\left(
1-\frac{|\mathbf{q}_j-\mathbf{q}'_l|^2}{6r_C^2}
\right),&\text{where}&|\mathbf{q}_j-\mathbf{q}'_l|\ll r_C.
\end{eqnarray}
Then, one can follow the same line of reasoning that we followed from Eq.~\eqref{eq:lambda}
to Eqs.~(\ref{eq:l_z}) and~(\ref{eq:l_x}) to solve the rest of integrations. Accordingly, the rates $\lambda_{x,z}$ of the DP model become:
\begin{eqnarray}
\lambda_z&=&
\frac{\Lambda_{\text{\tiny DP}}\,M}{8r_C^2\,m_0^2}
\,\int d\mathbf{Q}\,\left(|\psi_2(\mathbf{Q})|^2-|\psi_1(\mathbf{Q})|^2\right)\,
\sum_{j}m_j\,\mathbf{q}_j^2,
\\
\lambda_x&=&\lambda^{12}_{12}=\frac{\Lambda_{\text{\tiny DP}}}{2r_C^2\,m_0^2}
\left(
\int d\mathbf{Q}\,\psi_1(\mathbf{Q})\,\psi_2(\mathbf{Q})
\sum_{j}m_j \mathbf{q}_j
\right)^2.
\end{eqnarray}
with $\Lambda_{\text{\tiny DP}}=\frac{G\,m_0^2}{3\sqrt{2}\pi^{3/2}\,\hbar\,r_C}\simeq 7.39\times10^{-25}\,$s$^{-1}$. These rates have the same form as those of the CSL model, but with a different coupling constant, which is $10^{-16}$ times smaller than the CSL coupling constant, and therefore negligible. Accordingly, in the subsequent analysis we will report just the CSL values.
\vspace{2mm}

\section*{S6: Application to relevant physical systems.} We now provide quantitative estimation of these rates for some interesting physical systems.

\noindent{\it Hydrogen-like atoms.} 
For an atom that contains only one electron, we have: $\mathbf{D}_{12}=(m_e/e)\,\mathbf{d}_{12}$ with $m_e$ the mass and $e$ the charge of electron, and $\mathbf{d}_{12}$ the off-diagonal element of the dipole moment with typical values of a few Debye. In addition, $\sqrt{I_\alpha/m_e}$ has typical values of a few Bohr radius. Accordingly, we get: 
\begin{eqnarray}
\lambda_z\sim10^{-20}-10^{-18}\,\text{s}^{-1}; &&
\lambda_x\sim10^{-23}-10^{-21}\,\text{s}^{-1}.
\end{eqnarray}
For example, for the transition $2P\rightarrow1S$ of the hydrogen atom, where the emitted light is a K-level X-ray radiation~\cite{IUPAC}, %we get:
we find: $\lambda_z\simeq5.2\times10^{-19}\,$s$^{-1}$ and
$\lambda_x\simeq1.4\times10^{-22}\,$s$^{-1}$.

\noindent{\it Harmonic oscillator.}
We consider the two lowest states of a harmonic oscillator with mass $\mu$ and frequency $\omega_0$. Introducing these eigenstates into Eqs.~(\ref{eq:l_z}) and~(\ref{eq:l_x}) and performing the integration, we find:
\begin{eqnarray}
\lambda_x =
4\lambda_z = \frac{\Lambda}{2}\left(\frac{\mu\, x_0}{m_0\,r_C}\right)^2,
\end{eqnarray} 
with $x_0=\sqrt{\hbar/\mu\omega_0}$ the zero-point fluctuation amplitude. 

\noindent{\it Double-well potential.} We consider a system of mass $\mu$ moving in a symmetric double-well potential at low temperatures, where the meaningful eigenstates are the two lowest ones: $|\varepsilon_1\rangle=\frac{1}{\sqrt{2}}(|R\rangle+|L\rangle)$ and $|\varepsilon_2\rangle=\frac{1}{\sqrt{2}}(|R\rangle-|L\rangle)$.
The tunnelling frequency is $\omega_0=(\varepsilon_2-\varepsilon_1)/\hbar$. We denote the separation of minima by $q_0$. The states $|L\rangle$ and $|R\rangle$ are localized states at left and right minima, respectively. They can be transformed to each other by the displacement operator where the displacement distance is $q_0$, and $\langle L|R\rangle\simeq0$. Accordingly, using Eqs.~(\ref{eq:l_z}) and~(\ref{eq:l_x}) and performing the integration, we find:
\begin{eqnarray}
\lambda_z \simeq 0;&&\lambda_x \simeq \frac{\Lambda}{8}\left(\frac{\mu\, q_0}{m_0\,r_C}\right)^2.
\end{eqnarray}
For example, some internal motions of non-rigid molecules and complexes (e.g., the inversion motion in Ammonia~\cite{chi_book1} or hindered torsional rotation in X-Y-Y-X molecules~\cite{chi_book2}) can be effectively described by a double-well potential.  For these systems, we have $q_0\sim1-10\,$\AA$\;$ and $\mu\sim1-100\,$amu~\cite{chi_book1,chi_book2}. 
Accordingly, the order of magnitude of the strongest collapse rate for the internal motions of molecular systems is $\lambda_x\sim10^{-9}\,$s$^{-1}$. 

\end{widetext}

\end{document}